\documentclass[12pt]{article}
\usepackage{amsmath,amssymb,amscd}
\usepackage{listings}
\usepackage{caption}
\usepackage{slashed}
\usepackage{color}
\usepackage{ulem}

\usepackage{epstopdf}
\usepackage{epsfig}
\usepackage{listings}
\usepackage{caption}
\usepackage{amsmath}
\usepackage{breqn}
\newcommand{\beq}{\begin{equation}}
\newcommand{\eeq}{\end{equation}}
\newcommand{\nbea}{\begin{align*}}
\newcommand{\neea}{\end{align*}}
\newcommand{\nbeq}{\begin{equation*}}
\newcommand{\neeq}{\end{equation*}}
\usepackage{array}
\newcolumntype{M}[1]{>{\centering\arraybackslash}m{#1}}
\newcolumntype{N}{@{}m{0pt}@{}}

\numberwithin{equation}{section}
\begin{document}

\pagestyle{empty}

\baselineskip=21pt

\rightline{\footnotesize KCL-PH-TH/2017-26, CERN-TH/2017-101}
\vskip 0.75in

\begin{center}

{\large {\bf Anomaly-Free Models for Flavour Anomalies}}

\vskip 0.5in

 {\bf John~Ellis}$^{1,2}$
  {\bf Malcolm~Fairbairn}$^{1}$
and {\bf Patrick~Tunney}$^{1}$

%{John.Ellis@cern.ch}
%{malcolm.fairbairn@kcl.ac.uk}
%{patrick.tunney@kcl.ac.uk}

\vskip 0.5in

{\small {\it

$^1${Theoretical Particle Physics and Cosmology Group, Physics Department, \\
King's College London, London WC2R 2LS, UK}\\
\vspace{0.25cm}
$^2${Theoretical Physics Department, CERN, CH-1211 Geneva 23, Switzerland}\\
}}

\vskip 0.5in

{\bf Abstract}

\end{center}

\baselineskip=18pt \noindent

{\small
We explore the constraints imposed by the cancellation of triangle anomalies on models in which the flavour anomalies reported by LHCb and other experiments are due to an extra U(1)' gauge boson Z'. We assume universal and rational U(1)' charges for the first two generations of left-handed quarks and of right-handed up-type quarks but allow different charges for their third-generation counterparts. If the right-handed charges vanish, cancellation of the triangle anomalies requires all the quark U(1)' charges to vanish, if there are either no exotic fermions or there is only one Standard Model singlet dark matter (DM) fermion. There are non-trivial anomaly-free models with more than one such `dark' fermion, or with a single DM fermion if right-handed up-type quarks have non-zero U(1)' charges. In some of the latter models the U(1)' couplings of the first- and second-generation quarks all vanish, weakening the LHC Z' constraint, and in some other models the DM particle has purely axial couplings, weakening the direct DM scattering constraint. We also consider models in which anomalies are cancelled via extra vector-like leptons, showing how the prospective LHC Z' constraint may be weakened because the $Z^\prime \to \mu^+ \mu^-$ branching ratio is suppressed relative to other decay modes.
}

%%%%%%%%%%%%%%%%%%%%%%%%%%%%%%%%%%%%%%%%%%%%%%%%

\vskip 0.75in

%\leftline{ {%\small 
%May 2017}}

\newpage
\pagestyle{plain}

\section{Introduction}

The LHCb Collaboration and other experiments have reported a number of anomalies in semileptonic $B$ decays, including
apparent violations of $\mu - e$ universality in $B \to K^{(*)} \ell^+ \ell^-$ decays~\cite{Aaij:2017vbb,Bifani,Aaij1,Aaij2}, and apparent deviations from
the Standard Model (SM) predictions for the $P_5^\prime$ angular variable in $B \to K^* \ell^+ \ell^-$ decay~\cite{P5prime,Wehle} and
the $m_{\mu^+ \mu^-}$ distribution in $B_s \to \phi \mu^+ \mu^-$ decay~\cite{Aaij3}. These anomalies have reached a high
level of statistical significance. There are ongoing discussions about possible systematic effects and
the uncertainties in the SM calculations, so the jury is still out on the significances of these flavour anomalies~\cite{Ligeti}.
In the meantime, it is interesting to explore possible interpretations and look for other phenomenological signatures that might corroborate them~\cite{Sala,Capdevila,Altmannshofer,DAmico,Hiller,Geng,Ciuchini,Celis,Becirevic,Cai,Kamenik,DiChiara,Ghosh,Alok1,Alok2,Alonso,Wang,Greljo,Bonilla,Feruglio}.

In the wake of the recent LHCb report of $\mu - e$ non-universality in $B \to K^{*} \ell^+ \ell^-$ decay~\cite{Aaij:2017vbb,Bifani}, several
phenomenological analyses favour an anomalous non-SM contribution to the coefficient 
of the dimension-6 operator $O_9^\mu \equiv ({\bar s} \gamma_\mu P_L b)({\bar \mu} 
\gamma^\mu \mu)$~\cite{Capdevila,Altmannshofer,DAmico,Hiller,Geng,Ciuchini,Celis,Alok2} 
and do not exclude a smaller
non-SM contribution to the coefficient of $O_{10}^\mu \equiv ({\bar s} \gamma_\mu P_L b)({\bar \mu} \gamma^\mu \gamma_5 \mu)$ or $O_{9  \prime}^\mu \equiv ({\bar s} \gamma_\mu P_R b)({\bar \mu} 
\gamma^\mu \mu)$. 
A popular interpretation of this anomaly is that it is due to the exchange of a U(1)$^\prime$ gauge boson $Z'$
with non-universal couplings to both quarks and leptons~\cite{Sala,DAmico,Kamenik,DiChiara,Alok2,Alonso,Bonilla,someprevious,lmu-ltau}. 
It is possible to take a purely phenomenological attitude to this possibility, and not (yet) concern oneself about the 
theoretical consistency of such a $Z'$ model. However, any gauge theory should be free of triangle anomalies,
which in general need to be cancelled by fermions with masses comparable to that of the $Z'$. Thus, not only does 
this requirement have the potential
to constrain significantly the possible U(1)$^\prime$ couplings of both SM and non-SM particles, but it may
also suggest novel signatures that could confirm or disprove such a $Z'$ interpretation of the LHCb 
$B \to K^{*} \ell^+ \ell^-$ measurements and other flavour anomalies.

We explored recently the impact of the anomaly-cancellation requirement on simplified dark matter (DM)
models, assuming generation-{\it independent} U(1)$^\prime$ couplings to quarks and leptons~\cite{EFT1}. Rather
than take a top-down based on some postulated ultraviolet scenario, we
proposed some minimal benchmark models with desirable characteristics such as suppressed leptonic
couplings (so as to reduce the impact of unsuccessful LHC searches for massive $Z'$ bosons) or
axial coupling to quarks (so as to reduce the impact of direct searches for DM scattering). 

In this paper
we follow an analogous strategy for flavourful $Z'$ models with generation-{\it dependent} U(1)$^\prime$ 
couplings to quarks and leptons, treating DM as a possible optional extra. Motivated by the
long-standing discrepancy between experiment and the SM prediction for the anomalous magnetic
moment of the muon, $g_\mu - 2$, we also consider models with additional leptons $L$ that are
vector-like under the SM gauge group but might have parity-violating U(1)$^\prime$ couplings~\cite{Allanach}.

Since the LHCb anomaly could be explained by a left-handed flavour-changing quark coupling $\propto {\bar s} \gamma_\mu P_L b$, 
but not by a coupling $\propto {\bar s} \gamma_\mu P_R b$ alone, we initially follow~\cite{Allanach}
in assuming that the right-handed charge -1/3 quarks have vanishing U(1)$^\prime$
charges~\footnote{In contrast, the anomaly-free model proposed in~\cite{Bonilla} has a non-zero U(1)$^\prime$
charge for the $b_R$.}, but later include a scenario for anomaly cancellation with non-vanishing charges to right-handed charge -1/3 quarks in the Appendix.
In view of the strong upper limits on anomalous flavour-changing interactions of strange quarks,
we also assume~\cite{Allanach} that the first two generations of left-handed down-type quarks $d_L, s_L$ have identical U(1)$^\prime$ charges,
whereas the $b_L$ charge is different, opening the way to the suggested flavour-changing neutral interaction~\footnote{Models with similar $Z'$
couplings to left-handed charge -1/3 quarks were considered in~\cite{DAmico,Alonso},
the latter in the context of an anomaly-free horizontal symmetry motivated by the Pati-Salam~\cite{PS} model. See also~\cite{Alok2}, where the possibility of a weakly-coupled
light $Z'$ boson was also considered.}.
The reported $\mu - e$ non-universality and LEP constraints lead us also to assume that the electron has vanishing
U(1)$^\prime$ couplings~\cite{Allanach}, but we allow arbitrary U(1)$^\prime$ charges for the left- and 
right-handed $\mu$ and $\tau$,
generalizing the anomaly-free models discussed previously that assumed a $L_\mu - L_\tau$ charge in the lepton sector \cite{lmu-ltau}. The semileptonic $B$ decay data suggest that 
the couplings of the $\mu$ to the $Z'$ are predominantly vector-like, corresponding to dominance by the
$O_9^\mu$ operator over $O_{10}^\mu$, but we do not impose this restriction {\it a priori}.
We discuss anomaly-free models with muon couplings that are not completely vector-like, but have 
combinations of $O_9^\mu$ and $O_{10}^\mu$ that
are nevertheless consistent with a global fit the flavour anomalies, as seen in Figure \ref{fig:MoneyPlot}.

In the absence of U(1)$^\prime$ charges for the right-handed charge 2/3 quarks, we show that the anomaly-cancellation
conditions are so restrictive that there are no solutions with non-vanishing U(1)$^\prime$ charges for quarks. This is
also the case if we include a single dark sector particle.
However, we do find acceptable solutions if we allow for a second `dark' fermion, as illustrated in Table~\ref{tab:2DM}.

There are also solutions with a single DM particle if we allow non-vanishing
U(1)$^\prime$ charges for the $u_R, c_R$ (assumed to be equal) and $t_R$ (allowed be different). 
We have scanned for all possible triangle anomaly-free models
with charges that can be expressed in the form $p/q$ with $p, q \in [-4, 4]$ when we normalise the left-handed DM charge $Y^{\prime}_{\chi_L} = 1$.
Among these, 4 have vector-like $\mu$ couplings, 3 have no couplings to the first 2
generations of quarks, and 3 have axial couplings for the DM fermion (as required if it is a Majorana particle,
and which would suppress DM scattering by a relative velocity factor).
These models are all distinct, with the exception of a single model that combines a
vector-like $Z'$ coupling to the muon with an axial coupling to the DM particle.
We display in Table~\ref{tab:interesting} the U(1)$'$ charges for this model, two other models with vector-like muon couplings and one with vanishing $Z'$ couplings
for the first two generations, as benchmarks that illustrate the potential signatures of anomaly-free models
of flavourful $Z'$ bosons with DM.

We also explore models in which anomalies are cancelled by extra vector-like
leptons~\cite{Allanach}, exhibiting an example in which the LHC $Z' \to \mu^+ \mu^-$ signal is suppressed
because of a small $Z' \to \mu^+ \mu^-$ branching ratio.
Such a model may be able to explain the discrepancy between SM calculations and the experimental measurement of
the anomalous magnetic moment of the muon, $g_\mu - 2$~\cite{g-2}.

Finally, in the appendix, we note a construction which solves all the anomalies without the need for exotic fermions, but which requires a non-vanishing coupling to all types of quarks, including RH down-type quarks. In these models the coefficient of $O^{\mu}_{10}$ vanishes and there will be an admixture of the $O^{\mu}_9$ and $O^{\mu}_{9 \prime}$ operators.

\section{Scenarios for Anomaly Cancellation}

The following are the 6 anomaly-cancellation conditions to be considered:
\begin{enumerate}
\item[]{$(\rm a) \quad$ [SU(3)$_C^2$]$\times$[U(1)$^\prime$], which implies Tr[$\{\mathcal{T}^i,\mathcal{T}^j\}Y^\prime$] = 0,}
\item[]{$(\rm b) \quad$ [SU(2)$_W^2$]$\times$[U(1)$^\prime$], which implies Tr[$\{T^i,T^j\}Y^\prime$] = 0,}
\item[]{$(\rm c) \quad$ [U(1)$_Y^2$]$\times$[U(1)$^\prime$], which implies Tr[Y$^2 Y^\prime$] = 0,}
\item[]{$(\rm d) \quad$ [U(1)$_Y$]$\times$[U(1)$^{\prime ^2}$], which implies Tr[$Y Y^{\prime 2}$] =0,}
\item[]{$(\rm e) \quad$ [U(1)$^{\prime 3}$], which implies Tr[$Y^{\prime 3}$] =0,}
\item[]{$(\rm f) \quad$ Gauge-gravity, which implies Tr[$Y^\prime$] =0.}
\end{enumerate}
In general there there can be independent U(1)$'$ charges for each of the $3 \times 5 = 15$ multiplets of the Standard Model~\footnote{ Imposing
U(1)' invariance of Yukawa interactions and allowing for quark mixing would, in this case, require more than just the single Higgs multiplet of the SM.
Since this does not affect anomaly cancellation, we do not investigate this issue further here.},
which we label $q_{L,i}$, $u_{R,i}$, $d_{R,i}$, $l_{L,i}$ and $e_{R,i}$ where $i$ is a generation index,
as well as charges for any extra particles beyond the SM. However, as mentioned in the Introduction,
we make simplifying assumptions motivated by phenomenological considerations.

\noindent
{\it SM particles only, no $Z'$ couplings to electrons or right-handed quarks}

Motivated by the indication that the flavour anomalies originate in the U(1)$'$ couplings to left-handed
charge -1/3 quarks, initially we set the charges of all the right-handed quarks to zero: $Y'_{u_R,i} = 0 =Y'_{d_R,i}$,
though this is not mandated by the data. Motivated by the strong upper limits on non-SM
flavour-changing interactions between the first 2 generations of charge -1/3 quarks, we assume that the left-handed doublets in the first 2 generations have
identical charges $Y'_{q_L,1}=Y'_{q_L,2} \equiv Y'_q$~\cite{Allanach}. These differ from that of the third left-handed doublet $Y'_{q_L,3} \equiv Y'_t$,
making possible the desired flavour-changing $\bar{b} \gamma_\mu P_L s$ coupling. 

A complete discussion of the implications of
constraints on flavour-changing couplings, e.g., from $\Delta F = 2$ processes, is beyond the scope of this work,
since it would depend on the structures of the individual matrices that rotate the quark fields into the mass basis. 
Experimentally, only the combination entering into the CKM matrix is known, and the structures of the individual matrices
depend on details of the Higgs representations and Yukawa coupling matrices that are independent of the
anomaly cancellation conditions that we consider here. 

In order to avoid the experimental constraints
from LEP and other electroweak measurements~\cite{Allanach}, we also assume that the electron charges vanish: $Y'_{l_L,1}=0=Y'_{e_R,1}$.
However, we allow independent left- and right-handed couplings for the muon and tau. With these assumptions,
and in the absence of any particles beyond the SM, the anomaly-cancellation conditions become:
\begin{align}
2 (2Y'_q + Y'_t) &= 0 \label{eq:1anom1} \\
Y'_{\mu_L} + Y'_{\tau_L} + 3 Y'_{t} + 6 Y'_q &= 0 \label{eq:1anom2} \\
\frac{2}{3} \left( 3 Y'_{\mu_L} - 6 Y'_{\mu_R} + 3 Y'_{\tau_L} - 6 Y'_{\tau_R} + Y'_{t} + 2 Y'_q \right) &= 0 \label{eq:1anom3} \\
2 \left( - Y_{\mu_L}^{\prime \; 2} + Y_{\mu_R}^{\prime \; 2} - Y_{\tau_L}^{\prime \; 2} + Y_{\tau_R}^{\prime \; 2}+ Y_t^{\prime \; 2} + 2 Y_{q}^{\prime \; 2} \right) &= 0 \label{eq:1anom4} \\
-Y_{\mu_R}^{\prime \; 3} -Y_{\tau_R}^{\prime \; 3} + 2 \left( Y_{\mu_L}^{\prime \; 3} + Y_{\tau_L}^{\prime \; 3} \right) + 6 \left( Y_{t}^{\prime \; 3} + 2 Y_{q}^{\prime \; 3} \right) &= 0 \label{eq:1anom5} \\
-Y_{\mu_R}^{\prime} -Y_{\tau_R}^{\prime} + 2 \left( Y_{\mu_L}^{\prime} + Y_{\tau_L}^{\prime} \right) + 6 \left( Y_{t}^{\prime} + 2 Y_{q}^{\prime} \right) &= 0 \label{eq:1anom6}
\end{align}
Solving the conditions (\ref{eq:1anom1}, \ref{eq:1anom2}, \ref{eq:1anom3}) gives the relations 
\begin{align}
Y'_t &= -2 Y'_q \label{eq:elim1} \, , \\
Y'_{\tau_L} &= - Y'_{\mu_L} \label{eq:elim2} \, , \\
Y'_{\tau_R} &= - Y'_{\mu_R} \label{eq:elim3} \, .
\end{align}
Using these relations to solve the conditions (\ref{eq:1anom4}) then yields
\begin{align}
4 \left( -Y_{\mu_L}^{\prime \; 2} + Y_{\mu_R}^{\prime \; 2} + 3 Y_q^{\prime \; 2} \right) &=0 \, , \label{eq:1anom4sub}
\end{align}
which has rational solutions for any rational value of $Y'_q$, since any odd number can be written as the difference between
two squares. Equation (\ref{eq:1anom4sub}) implies that if $Y'_q \ne 0$ the muon could not have vector-like U(1)$'$ couplings as suggested by the data, but this is a
moot point in this scenario, since (\ref{eq:1anom5}) yields
\begin{align}
- 36 Y_q^{\prime \; 3} &= 0 \, . \label{eq:1anom5sub}
\end{align}
Hence $Y'_q = 0$, which in turn implies via (\ref{eq:elim1}) that
$Y'_t$ also vanishes and the $Z'$ decouples from quarks~\footnote{The possibility that the $Z'$ couples only
to $t$ quarks and that the LHCb flavour anomalies are loop-induced was considered in~\cite{Kamenik}.}. 
Therefore, we must relax the assumptions made above if the flavour anomalies are to
be explained by the exchange of a $Z'$ boson.\\

\noindent
{\it Including one or two `dark' particles}

Adding a single DM particle with vanishing SM couplings does not remedy the situation, 
as none of the conditions (\ref{eq:elim1}, \ref{eq:elim2}, \ref{eq:elim3}) are affected, and
condition (\ref{eq:1anom6}) implies $\mathrm{Tr}_{\mathrm{BSM}} [Y^{\prime}] = 0$. Hence, if
there is a single DM particle it must have vector-like U(1)$'$ couplings, and would not change the
fatal condition (\ref{eq:1anom5sub}).

\begin{figure}[h!]

\centering

\includegraphics[scale=0.55]{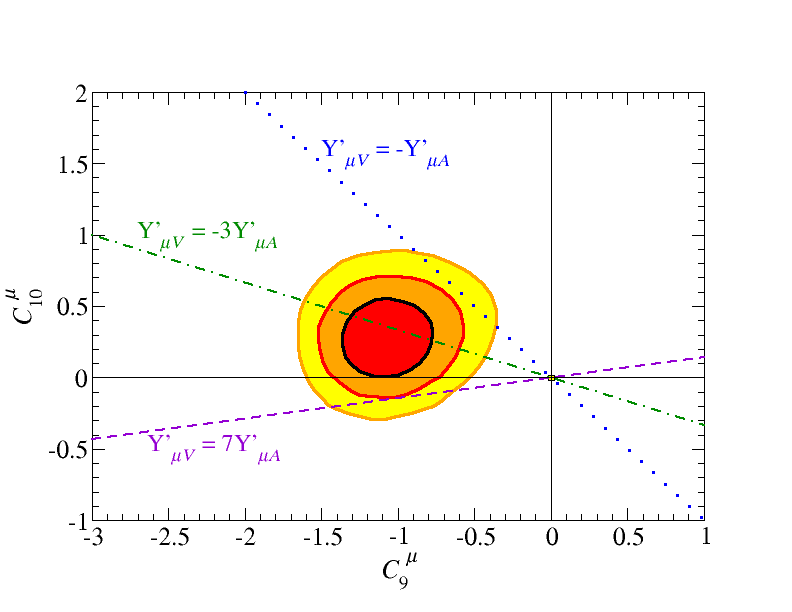}

\caption{\it Regions of the operator coefficients $(C_9^\mu, C_{10}^\mu)$ preferred in the global analysis of flavour anomalies in~\protect\cite{Capdevila} ,
which includes all relevant flavour observables including, e.g., the branching fraction for $B_s \to \mu \mu$. 
We superimpose the predictions of models with $Y'_{\mu_V}/Y'_{\mu_A} = -3$ as in Table~\protect\ref{tab:2DM}
(green dot-dashed line), $Y'_{\mu_V}/Y'_{\mu_A} = 7$ (purple dashed line), e.g., model (D) in Table~\protect\ref{tab:interesting}  and $Y'_{\mu_V}/Y'_{\mu_A} = -1$ (blue dotted line) as in model (E) in Table~\protect\ref{tab:interesting}.
$C_{10}^\mu = 0$ in models (A, B) and (C) in Table~\protect\ref{tab:interesting} and in the model in Table~\protect\ref{tab:suppress}.}
\label{fig:MoneyPlot}
\end{figure}

The next simplest possibility has two SM-singlet `dark' fermions, $A$ and $B$, in which
case we have, in addition to (\ref{eq:elim1},\ref{eq:elim2}) and (\ref{eq:elim3}) the
condition that
\begin{equation}
Y'_{A_L} = Y'_{A_R} -Y'_{B_L} + Y'_{B_R} \label{eq:anom6DM2}
\end{equation}
and the remaining anomaly conditions to solve are (\ref{eq:1anom4sub}) and
\begin{align}
%4 \left( -Y_{\mu,L}^{\prime \; 2} + Y_{\mu,R}^{\prime \; 2} + 3 Y_q^{\prime \; 2} \right) &=0 \, ,  \\
-36 Y^{\prime \; 3}_{q} + 3 (Y'_{B_L} - Y'_{A_R}) (Y'_{B_L} - Y'_{B_R}) (Y'_{A_R} + Y'_{B_R}) &=0
\label{anom5DM2}
\end{align}
instead of (\ref{eq:1anom5sub})
As in the single DM particle case, the condition (\ref{eq:1anom4sub}) implies that the muon cannot have vector-like U(1)$'$ couplings.
Normalizing $Y'_{A_R} = 1$ and restricting our attention to anomaly-free models with U(1)$'$ charges
that can be expressed as $p/q$ with $p, q \in [-4, 4]$, we find several solutions with $Y'_{\mu_L}/Y'_{\mu_R} = 2$~\footnote{We
also find models with vanishing $Y'_{A_R}$ and the same ratio $Y'_{\mu_L}/Y'_{\mu_R} = 2$.}. 
In these models the ratio of the vector-like and axial muon couplings $Y'_{\mu_V}/Y'_{\mu_A} = -3$, which
may be consistent with the ratio of $O_9^\mu$ and $O_{10}^\mu$ coefficients $C_9^\mu, C_{10}^\mu$ allowed by
the analyses in~\cite{Capdevila,Altmannshofer,DAmico,Hiller,Geng,Ciuchini,Celis,Alok2}, as indicated in Fig.~\ref{fig:MoneyPlot}.
We see that the green dot-dashed line corresponding to models with $Y'_{\mu_V}/Y'_{\mu_A} = -3$ traverses the region of the 
$(C_9^\mu, C_{10}^\mu)$ plane preferred in the analysis of~\cite{Capdevila} at the 1-$\sigma$ level.
These models have the same U(1)$'$ charges for the SM particles but different values for Y$'_{A_R, B_{L,R}}
= 0, \pm 1/3, \pm 4/3$. None of these solutions has a DM candidate with a purely axial U(1)$'$
coupling, though we cannot exclude the possibility that the SM-singlet fermions might mix in such a way that the lighter mass eigenstate does have an
axial coupling. The U(1)$'$ charges of one representative model are shown in Table~\ref{tab:2DM}.

The identification of the lighter SM-singlet fermion mass eigenstate depends on details of the mixing in the dark sector
that we do not discuss here. Various experimental constraints should be considered for this fermion to be a realistic DM candidate: 
the correct thermal relic density should be obtained, 
the cross-sections for scattering on nuclei should be below the sensitivities of current direct detection experiments, 
and LHC and indirect detection bounds should be taken into account where appropriate. 
Anomaly cancellation constrains only the $Y'$ charges but not the overall magnitude $g$ of the gauge coupling, 
which could be fixed by the requiring the observed abundance of dark matter. When combined with the $Y'$ charges
and mixing patterns in specific models, predictions for the LHC and dark matter experiments could be made, but such a study
lies beyond the scope of this work.

\vspace{0.5cm}
\begin{table}[h]
{\small
\begin{center}
{\bf Models with only left-handed quark couplings and two dark fermions}\\
\vspace{4mm}
\begin{tabular}
{ | c | c | c | c | c | c | c | c | c | c |}
\hline
Y$'_{q_L}$ & Y$'_{t_L}$ & Y$'_{\mu_L}$ & Y$'_{\mu_R}$ & Y$'_{\tau_L}$ & Y$'_{\tau_R}$ & Y$'_{A_L}$ & Y$'_{A_R}$ & Y$'_{B_L}$ & Y$'_{B_R}$\\
%\vspace{1mm}
\hline
%\hline
%\multicolumn{11}{| c |}{ Vector-like $\mu$ coupling and axial DM coupling} \\
\hline
%(A) & 0 & 1 & 1 & 0 & -2 & -2 & -1 & -2 & 1 & -1 & ? & ? \\
%\hline
%\hline
%\multicolumn{11}{| c |}{ Vector-like $\mu$ couplings} \\
%\HLINE
1/3 & -2/3 & 2/3 & 1/3 & -2/3 & -1/3 & 0 & 1 & -1/3 & -4/3 \\
\hline
\end{tabular}
\end{center}
}
\caption{\it The U(1)$'$ charges in a benchmark model with two SM-singlet `dark' particles $A, B$
that have couplings only for left-handed quarks and a muon coupling that is dominantly vector-like:
$Y'_{\mu_V}/Y'_{\mu_A} = -3$.}
\label{tab:2DM}
\end{table}

\noindent
{\it Including couplings for right-handed charge 2/3 quarks}

As an alternative way to relax our initial assumptions, we allow next for non-vanishing U(1)$'$ charges for
the right-handed charge 2/3 quarks $u_{R,i}$. We recall that the flavour anomalies apparently originate from
left-handed $b$ and $s$ quarks,
however there is no reason to forbid U(1)$'$ couplings to right-handed charge 2/3 quarks. For the moment, we assume that the charges of the RH charge -1/3 quarks vanish, but relax this assumption in the Appendix. We assume that the U(1)$'$ charges of the first two generations are identical, i.e., 
$Y'_{u_R,1} = Y'_{u_R,2}$, again with the motivation of suppressing flavour-changing neutral interactions.

In the absence of non-SM fermions, one can readily solve the anomaly conditions (a, b, c) and (f)
above, and the conditions (d) and (e) then take the following forms:
\begin{align}
Y^{\prime \; 2}_{l_L,2} - Y^{\prime \; 2}_{e_R,2} - 3 Y^{\prime \; 2}_{q_L,1} + 6 Y^{\prime \; 2}_{u_R,1} &= 0 \, , \label{eq:2anom4sub} \\
-2 Y^{\prime \; 3}_{q_L,1} + Y^{\prime \; 3}_{u_R,1} & = 0 \, . \label{eq:2anom5sub}
\end{align}
These conditions clearly have non-trivial solutions, but (\ref{eq:2anom5sub}) does not admit
rational values for both $Y^{\prime}_{q_L,1}$ and  $Y^{\prime}_{u_R,1}$. All the unification scenarios
known to us have rational values for U(1)$'$ charges, so these solutions are not acceptable.\\

\noindent
{\it Including couplings for right-handed charge 2/3 quarks and a DM particle}

We are therefore led to consider adding a single DM fermion $\chi$ with charges $Y'_{\chi_L}$ and $Y'_{\chi_R}$.
Normalizing $Y'_{\chi_L} = 1$, the anomaly conditions (a, b, c, d) and (f) yield the following expressions for the other charges
\begin{align}
Y'_{t_L} &= -\frac{-2 Y'_{q_L} \left(-3 Y'_{\mu _L}+8 Y'_{q_R}+4 Y'_{\mu
   _R}\right)+\left(Y'\right)_{q_L}^2+\left(Y'\right)_{\mu _L}^2+6
   \left(Y'\right)_{q_R}^2-\left(Y'\right)_{\mu _R}^2}{2 Y'_{q_L}+3 Y'_{\mu _L}-8 Y'_{q_R}-4 Y'_{\mu
   _R}} \label{eq:YTL} \, , \\
Y'_{t_R} &= \frac{2 \left(Y'_{q_R} \left(4 Y'_{\mu _R}-3 Y'_{\mu _L}\right)-2 Y'_{q_L} Y'_{q_R}+3
   \left(Y'\right)_{q_L}^2-\left(Y'\right)_{\mu _L}^2+2 \left(Y'\right)_{q_R}^2+\left(Y'\right)_{\mu
   _R}^2\right)}{2 Y'_{q_L}+3 Y'_{\mu _L}-8 Y'_{q_R}-4 Y'_{\mu _R}} \label{eq:YTR} \, , \\
Y'_{\tau, L} &= \frac{8 Y'_{\mu _L} Y'_{q_R}-2 Y'_{q_L} Y'_{\mu _L}-9 \left(Y'\right)_{q_L}^2+Y'_{\mu _R} \left(4
   Y'_{\mu _L}-3 Y'_{\mu _R}\right)+18 \left(Y'\right)_{q_R}^2}{2 Y'_{q_L}+3 Y'_{\mu _L}-8 Y'_{q_R}-4
   Y'_{\mu _R}} \, , \label{eq:YTAL} \\
Y'_{\tau, R} &= \frac{-2 Y'_{q_L} Y'_{\mu _R}-12 \left(Y'\right)_{q_L}^2+Y'_{\mu _L} \left(4 Y'_{\mu _L}-3 Y'_{\mu
   _R}\right)+8 Y'_{q_R} Y'_{\mu _R}+24 \left(Y'\right)_{q_R}^2}{2 Y'_{q_L}+3 Y'_{\mu _L}-8 Y'_{q_R}-4
   Y'_{\mu _R}} \, , \label{eq:YTAR} \\
Y'_{\chi_R} &= \frac{-6 \left(Y'\right)_{q_L}^2+2 Y'_{q_L}+2 \left(Y'\right)_{\mu _L}^2+3 Y'_{\mu _L}+12
   \left(Y'\right)_{q_R}^2-8 Y'_{q_R}-2 \left(Y'\right)_{\mu _R}^2-4 Y'_{\mu _R}}{2 Y'_{q_L}+3 Y'_{\mu
   _L}-8 Y'_{q_R}-4 Y'_{\mu _R}} \, , \label{eq:YXR}
\end{align}
when $2 Y'_{q_L}+3 Y'_{\mu_L}-8 Y'_{q_R}-4 Y'_{\mu _R} \neq 0 $~\footnote{There are no
`interesting' solutions with vector-like muon couplings, vanishing first- and second-generation couplings
or axial DM couplings when $2 Y'_{q_L}+3 Y'_{\mu_L}-8 Y'_{q_R}-4 Y'_{\mu _R} = 0 $.}.
At this stage $Y'_{q_L}$, $Y'_{q_R}$, $Y'_{\mu _L}$ and $Y'_{\mu _R}$ are undetermined, but we have not yet applied the
anomaly condition (e), which yields an additional constraint that is algebraically complicated and unrevealing.
Scanning over the four undetermined charges, we find a set of solutions with $Y'_{q_L}=Y'_{q_R}=0$
(which would suppress $Z'$ production at the LHC and direct DM scattering) and either 
$Y'_{\mu _L}=Y'_{\mu _R}$ (as favoured by the data) or $Y'_{\mu _L}= -Y'_{\mu _R}$. However, 
these solutions also have vanishing couplings for the third-generation quarks, i.e., $Y'_{t_L} = Y'_{t_R} = 0$, so all 
the quark charges vanish.

We are therefore forced to make a `Solomonic choice' between models with vector-like couplings to
muons, i.e.,  $Y'_{\mu _L}=Y'_{\mu _R}$, and those with vanishing couplings to first- and second-generation quarks.
Scanning over rational values of U(1)$'$ that can be expressed in the form $p/q: p, q \in [-4, 4]$, we find 4 models with vector-like muon couplings
and 3 that have vanishing first- and second-generation quark couplings. One of the models with vector-like muon couplings 
also has a DM particle with a purely axial U(1)$'$ coupling that could be a Majorana particle. The U(1)$'$ charges of 
this model (A) are listed in Table~\ref{tab:interesting}, along with the corresponding charges for some other models that 
may serve as interesting benchmarks. 
The charges in the second and third rows are for models (B, C) with vector-like muon
couplings but non-axial DM couplings, and the charges in the bottom two rows are for a model (D,E) with vanishing first- and
second-generation quark couplings and a mixture of vector and axial couplings to the muon.

\vspace{0.5cm}
\begin{table}[h]
{\small
\begin{center}
{\bf Models with right-handed charge 2/3 quark couplings and one DM fermion}\\
\vspace{4mm}
\begin{tabular}
{ | c | c | c | c | c | c | c | c | c | c | c | }
\hline
 & Y$'_{q_L}$ & Y$'_{q_R}$ & Y$'_{t_L}$ & Y$'_{t_R}$ & Y$'_{\mu_L}$ & Y$'_{\mu_R}$ & Y$'_{\tau_L}$ & Y$'_{\tau_R}$ & Y$'_{\chi_L}$ & Y$'_{\chi_R}$ \\
%\vspace{1mm}
\hline
\hline
\multicolumn{11}{| c |}{ Vector-like $\mu$ coupling and axial DM coupling} \\
\hline
(A) & 0 & 1 & 1 & 0 & -2 & -2 & -1 & -2 & 1 & -1 \\
\hline
\hline
\multicolumn{11}{| c |}{ Vector-like $\mu$ couplings} \\
\hline
(B) & 1/3 & 1/3 & -1/3 & 0 & -1 & -1 & 0 & -1/3 & 1 & 1/3 \\
(C) & 1/2 & 0 & -1/2 & 1 & -1/2 & -1/2 & -1 & -3/2 & 1 & 0 \\
\hline
\hline
\multicolumn{11}{| c |}{ No first- and second-generation couplings} \\
\hline
(D) & 0 & 0 & 1/2 & 1 & -3/2 & -2 & 0 & 0 & 1 & 0 \\
(E) & 0 &	0 & 1/2 & 1 & -3/2 & 0 & 0 & -2 & 1 & 0 \\
\hline
\end{tabular}
\end{center}
}
\caption{\it The U(1)$'$ charges in some benchmark models with couplings for right-handed quarks and a single
dark matter particle that have interesting properties:
(A) vector-like $\mu$ coupling and axial DM coupling, (B, C) vector-like $\mu$ coupling,
(D) no first- and second-generation couplings and relatively small axial-vector $\mu$ coupling: $Y'_{\mu_V}/Y'_{\mu_A} = 7$., (E) no first- and second-generation couplings and $Y'_{\mu_V}/Y'_{\mu_A} = -1$}
\label{tab:interesting}
\end{table}

In models such as (D,E), the $Z'$ production mechanisms via first- and second-generation ${\bar q} q$ annihilations 
that are usually dominant at  the LHC are suppressed,
and the constraint on the $Z'$ mass coming from production via ${\bar b} s + {\bar s} b$ collisions is much weaker~\cite{DiChiara}.
Moreover, the constraint from searches for direct DM scattering on nuclei is greatly weakened. 

Although the $Z'$ coupling
to muons is not purely vectorial in model (D), the ratio of the vector and axial muon couplings is 7
in this model, so the axial coupling might be acceptably 
small. As seen in Fig.~\ref{fig:MoneyPlot}, the data allow a small axial/vector ratio, although they prefer the opposite relative sign.
Models with $Y'_{\mu_V}/Y'_{\mu_A} = 7$ (dashed purple line) are compatible with the region of the $(C_9^\mu, C_{10}^\mu)$ preferred in the analysis
of~\cite{Capdevila} at the 2-$\sigma$ level.

Model (E) shares the property of having no coupling to the first two generations of quarks but has a different mixture of axial/vector coupling, $Y'_{\mu_V}/Y'_{\mu_A} = -1$ since it has a purely left-handed muon coupling. This is compatible with the fit shown in Fig. \ref{fig:MoneyPlot} at the $3 \, \sigma$ level.

We also found a model (not shown) with vanishing first- and second-generation quark couplings, 
but with a muon coupling that is either purely right-handed, which is disfavoured by the data~\footnote{We also find models
with $Y'_{\chi_L} = 0$ but with $Y'_{\chi_R}$ non-vanishing, in which $Y'_{\mu_V}/Y'_{\mu_A} = 7$ and 0, as well as $-1$, which appears to compatible with the data at the 3-$\sigma$ level.}.\\

\noindent
{\it Including a vector-like lepton}

Finally, we consider a scenario proposed in~\cite{Allanach} in which the SM particles are not supplemented by DM, but by extra
leptons, a vector-like doublet $(\nu', \ell')$ and a vector-like singlet $\mu'$. We assume that only the left-handed quarks have
non-zero U(1)$'$ charges, with those for the first two generations being the same. We also assume that the $Z'$ coupling of
the muon is purely vectorial. The left-handed components of the doublet and the
right-handed component of the singlet are assumed to have identical values of $Y'$, but the U(1)$'$ charges of the right-handed
doublet and left-handed singlet are free {\it a priori}. Thus the free parameters of the model are $Y'_{q_L}, Y'_{t_L},
Y'_{\mu_L} = Y'_{\mu_R}, Y'_{\tau_{L,R}}, Y'_{\ell'_L} = Y'_{\mu'_R}, Y'_{\ell'_R}$ and $Y'_{\mu'_L}$.

\vspace{0.5cm}
\begin{table}[h!]
{\small
\begin{center}
{\bf Model with extra vector-like leptons}\\
\vspace{4mm}
\begin{tabular}
{ | c | c | c | c | c | c | c | c | c |}
\hline
Y$'_{q_L}$ & Y$'_{t_L}$ & Y$'_{\mu_L}$ & Y$'_{\mu_R}$ & Y$'_{\tau_L}$ & Y$'_{\tau_R}$ & Y$'_{\ell'_L} = $ Y$'_{\mu'_R}$ & Y$'_{\ell'_R}$ & Y$'_{\mu'_L}$  \\
%\vspace{1mm}
\hline
\hline
1 & -2 & 1 & 1 & 4 & 3 & -4 & 1 & 0 \\
\hline
\end{tabular}
\end{center}
}
\caption{\it The U(1)$'$ charges in a model with extra vector-like leptons $\ell', \mu'$ and a vector-like $Z'$
muon coupling in which the branching ratio for $Z' \to \mu^+ \mu^-$ is suppressed.}
\label{tab:suppress}
\end{table}

We have also scanned rational values of these free parameters that can be expressed in the form $p/q: p, q \in [-4, 4]$. 
Since one of the objectives of~\cite{Allanach} is to explain the discrepancy between SM calculations of $g_\mu -2$
and the experimental measurement~\cite{g-2}, via a contribution $\propto 1/M_{Z^{'2}}$, it is desirable to focus on solutions in which the
LHC $Z' \to \mu^+ \mu^-$ signal is suppressed. Since the U(1)$'$ charges of the first- and second-generation quarks are
non-vanishing, the only way to suppress the prospective LHC signal is to suppress the $Z' \to \mu^+ \mu^-$ branching ratio.
We have found several models in which the combined branching ratios for other decays exceed that for 
$Z' \to \mu^+ \mu^-$ by more than an order of magnitude. Table~\ref{tab:suppress} displays the model in
which the branching ratio for $Z' \to \mu^+ \mu^-$ is most suppressed by the U(1)$'$ charges and multiplicities of states,
namely to 3/130, assuming that the masses of the extra leptons can be neglected, as is the case if all the
fermions are much lighter than $M_{Z^\prime}/2$. Since $M_{Z'}$ may be in the TeV range,
this is compatible with the lower limits on the masses of
vector-like leptons given by the Particle Data Group~\cite{PDG}, which are $\sim 100$~GeV, 
and also with model-dependent recasts of LHC searches~\cite{FSV}, which yield limits $\sim$ a few hundred GeV.

\section{Summary and Conclusions}

We have explored in this paper the constraints on $Z'$ interpretations of the flavour anomalies in $B \to K^{(*)} \ell^+ \ell^-$
decays imposed by the cancellation of triangle anomalies, namely the conditions (a) to (f) stated at the beginning of Section~2.
We find many models that have not been discussed previously in the literature, and have novel experimental
signatures involving new particles and/or non-trivial combinations of the operators $O_9^\mu$ and $O_{10}^\mu$ 
that are consistent with the reported flavour anomalies.

Motivated by the
observed pattern of flavour anomalies, we considered initially models in which the $Z'$ has quark couplings that are purely left-handed
(universal for the first 2 generations, non-universal for the third), and has no electron coupling. In this case we find no non-trivial
solution of the anomaly-cancellation conditions in the absence of non-SM particles, and so are led to introduce `dark' fermions 
without SM couplings. In the case of a single DM particle, there is again no non-trivial solution, but we do find
solutions with 2 `dark' fermions. In none of these does the $Z'$ have a purely vector-like muon coupling, but we find a
class of solutions in which $Y'_{\mu_V}/Y'_{\mu_A} = -3$, a ratio that is compatible with the data at the 1-$\sigma$ level
as seen in Fig.~\ref{fig:MoneyPlot}. Examples of these solutions are shown in Table~\ref{tab:2DM}.

We then considered models in which the $Z'$ couples to right-handed charge 2/3 quarks, a possibility that is allowed by the data.
In the absence of a dark sector we find no solution of the anomaly-cancellation conditions with rational charges, but we do find a
number of interesting solutions in the presence of a DM fermion, and we show some examples in Table~\ref{tab:interesting}. Some of these
have vector-like muon couplings - models (A), (B) and (C) - and in one of these the DM particle has a purely axial $Z'$ coupling -
model (A). In model (D,E) there are no $Z'$ couplings to first- and second-generation quarks, so production at the LHC is suppressed
and the experimental constraints on the dark mass scale are correspondingly reduced. Model (D) is one of a class of models in which
$Y'_{\mu_V}/Y'_{\mu_A} = 7$, a ratio that appears compatible with the data at the 2-$\sigma$ level, as also seen in Fig.~\ref{fig:MoneyPlot}. 
Model (E) predicts instead $Y'_{\mu_V}/Y'_{\mu_A} = -1$, originating from a pure left-handed muon coupling, and is compatible with the data at the 3-$\sigma$ level.
We have also considered models in which the triangle anomalies are cancelled by vector-like leptons, exhibiting in Table~\ref{tab:suppress}
a model with a vector-like $Z'$ muon coupling in which the branching ratio for $Z' \to \mu^+ \mu^-$ is maximally suppressed.

These examples illustrate that anomaly cancellation is a powerful requirement that could have interesting
phenomenological consequences linking flavour anomalies to other observables. 
Anomaly cancellation requires some extension of the SM spectrum to include, e.g., a dark sector or a
vector-like lepton. Moreover, either the dark sector should more than just a single DM particle, or some quarks should have
right-handed couplings to the $Z'$ boson. Additionally, we find several classes of models in which $Y'_{\mu_V}/Y'_{\mu_A} \ne 0$
in a way that is compatible with the present data but could be explored in the future. Finally, we have shown that it is possible
to cancel the triangle anomalies using vector-like leptons in such a way as to suppress the LHC $Z' \to \mu^+ \mu^-$ signal,
potentially facilitating an explanation of the anomaly in $g_\mu - 2$.

\section*{Acknowledgements}

The work of JE and MF was supported partly by the STFC Grant ST/L000326/1. 
MF and PT are funded by the European Research Council under the European Union's Horizon 2020 programme
(ERC Grant Agreement no.648680 DARKHORIZONS).  We thank Ben Allanach for discussions.

\section*{Note added}

After the appearance of this manuscript an updated calculation of the SM prediction for the $B_s$ mass difference \cite{DiLuzio:2017fdq} appeared which strongly constrains new physics scenarios via their contribution to the $\bar{b}s + \bar{s} b$ coupling. The updated SM prediction is around $1.8 \, \sigma$ above the experimental measurement, meaning that new physics scenarios that give a positive contribution to $\Delta M_s$ are strongly constrained at the $2 \, \sigma$ confidence level. All models described here, with the exception of the ones in the Appendix, give only a positive contribution to $\Delta M_s$, assuming real couplings, since we do not couple to RH down-type quarks.

We have examined the impact of these constraints for the model in Table \ref{tab:2DM} (denoted ``model 1'') and model D in Table \ref{tab:interesting} (denoted ``model 2'') and found that for a high mass (TeV range) Z', model 1 is in significant tension with Bs mixing and dilepton searches whereas model 2D is viable due to it's vanishing coupling to first and second generation quarks.

More specifically, for model 1 to fit the combined analysis of $b \rightarrow s l^+ l^-$ flavour anomalies shown in Fig. \ref{fig:MoneyPlot} at the $3 \, \sigma$ level, while satisfying the new $B_s$ mass mixing bound at $3 \, \sigma$ requires $M_{Z'} < 19.0$ TeV for a fixed gauge coupling $g=6$. This coupling sets the muon coupling close to the non-perturbative limit, which allows for the highest mass Z' that can explain the flavour anomalies while satisfying $B_s$ mixing \cite{DiLuzio:2017fdq}. This Z' mass is beyond the reach of the latest ATLAS dilepton resonance search \cite{Aaboud:2017buh}, but we expect that the published limits on the non-resonant region to rule out model 1 since our couplings are close to the non perturbative limit (specifically for the limit on the LL const. interaction from ATLAS, rules out Mz' < 24 TeV for our model 1).

Model 2D can fit the flavour anomalies at the $2 \, \sigma$ level and satisfy the updated $B_s$ bound at the $2 \, \sigma$ level for $M_{Z'} < 2.8$ TeV with a fixed gauge coupling of 1.5. Note we do not restrict to CKM mixing here, our mixing factor which the $\bar{b}s$ coupling is proportional to is 0.006 at this point in parameter space ($M_{Z'} = 2.8$ TeV). The width of the Z' boson here is 32\%, ATLAS \cite{Aaboud:2017buh} require $\sigma B A < 2.6 \times 10^{-4}$ fb, where $\sigma$ is the production cross section, B the branching ratio into muons and A the detector acceptance.

We leave a more detailed study of whether this and the other models are ruled out to a later study, but note that the $Z'$ production via $\bar{b}s + \bar{s} b$ annihilations is suppressed by a small mixing factor $\sim$ 0.006. This is to be compared to many other models in the literature where the mixing is $\sim V_{ts} V_{tb} \sim 0.04$.

\section*{Appendix: Non-vanishing couplings to right-handed down-type quarks}

We now investigate the effect of anomaly cancellation with couplings to all quark fields, including charges for the right-handed down-type quarks. As before, we fix the charges of the first two quark generations to be equal. However in order to restrict the number of unknowns we take a purely vectorial coupling, $Y_{\mu,L }'=Y_{\mu,R }'$ such that $C^{\mu}_{10} = 0$. In this case the anomaly cancellation conditions read

\begin{dgroup}
\begin{dmath}
0 = 2 Y_3'-Y_b'-2 Y_d'+4 Y_q'-Y_t'-2 Y_u'  \label{eq:AllqAnom1}
\end{dmath}
\begin{dmath}
0 = 3 Y_3'+6 Y_q'+Y_{\mu }'+Y_{\tau ,L}'  \label{eq:AllqAnom2} 
\end{dmath}
\begin{dmath}
0 = \frac{2}{3} \left(Y_3'-2 Y_b'-4 Y_d'+2 Y_q'-8 Y_t'-16 Y_u'-3 Y_{\mu
   }'+3 Y_{\tau ,L}'-6 Y_{\tau ,R}'\right)  \label{eq:AllqAnom3}
   \end{dmath}
\begin{dmath}
0 = \left(Y_3'\right){}^2+\left(Y_b'\right){}^2+2
   \left(Y_d'\right){}^2+2 \left(Y_q'\right){}^2 
   - 2  \left(Y_t'\right){}^2-4 \left(Y_u'\right){}^2 -\left(Y_{\tau
   ,L}'\right){}^2+\left(Y_{\tau ,R}'\right){}^2   \label{eq:AllqAnom4}
   \end{dmath}
\begin{dmath}
0 = 6 \left(Y_3'\right){}^3-3 \left(Y_b'\right){}^3-6
   \left(Y_d'\right){}^3+12 \left(Y_q'\right){}^3-3
   \left(Y_t'\right){}^3-6 \left(Y_u'\right){}^3+\left(Y_{\mu
   }'\right){}^3+2 \left(Y_{\tau ,L}'\right){}^3-\left(Y_{\tau
   ,R}'\right){}^3 \label{eq:AllqAnom5} 
   \end{dmath}
\begin{dmath}
0 = 6 Y_3'-3 Y_b'-6 Y_d'+12 Y_q'-3 Y_t'-6 Y_u'+Y_{\mu }'+2 Y_{\tau
   ,L}'-Y_{\tau ,R}' \label{eq:AllqAnom6}
\end{dmath}
\end{dgroup}
with $Y_3'= Y_{q,L,3}'$ and $Y_t' = Y'_{u,R,3}$ and $ Y_q' = Y_{q,L,1/2}'$ and $Y_u' = Y'_{u,R,1/2}$  etc. Also $Y_{\mu }' = Y_{\mu,L }'=Y_{\mu,R }'$.

With all these charges present, the anomaly cancellation can be solved completely algebraically. We have the following constraints in terms of the unconstrained charges $Y'_q$, $Y'_u$ and $Y'_d$:

\begin{dgroup}
\begin{dmath}
Y'_3 = -\frac{2 \left(Y_d'\right){}^2 Y_q'+\left(Y_d'\right){}^3-4 Y_q'  \left(Y_u'\right){}^2+\left(Y_u'\right){}^3}{\left(Y_d'\right){}^2+\left(Y_q'\right){}^2-2 \left(Y_u'\right){}^2} \label{eq:AllqY3} 
\end{dmath}
\begin{dmath}
Y'_{\tau , L} = 
\frac{1}{6} \left(\frac{9 Y_u' \left(\left(Y_d'\right){}^2+\left(Y_q'\right){}^2\right)+18 \left(\left(Y_d'\right){}^3-2 \left(Y_q'\right){}^3\right)}{\left(Y_d'\right){}^2+\left(Y_q'\right){}^2-2 \left(Y_u'\right){}^2}+\frac{24 \left(Y_u'\right){}^2 \left(\left(Y_d'\right){}^2+\left(Y_q'\right){}^2\right)+24 Y_u' \left(\left(Y_d'\right){}^3-2 \left(Y_q'\right){}^3\right)-6 \left(\left(Y_d'\right){}^2+\left(Y_q'\right){}^2\right){}^2}{\left(Y_d'\right){}^3-2 \left(Y_q'\right){}^3+\left(Y_u'\right){}^3}+8 Y_d'-4 Y_q'-Y_u'\right) 
\label{eq:AllqYTAL} 
\end{dmath}
\begin{dmath}
Y'_{\tau , R} = 
\frac{3 Y_u' \left(\left(Y_d'\right){}^2+\left(Y_q'\right){}^2\right)+6 \left(\left(Y_d'\right){}^3-2
   \left(Y_q'\right){}^3\right)}{\left(Y_d'\right){}^2+\left(Y_q'\right){}^2-2 \left(Y_u'\right){}^2}+\frac{4 \left(Y_u'\right){}^2
   \left(\left(Y_d'\right){}^2+\left(Y_q'\right){}^2\right)+4 Y_u' \left(\left(Y_d'\right){}^3-2
   \left(Y_q'\right){}^3\right)-\left(\left(Y_d'\right){}^2+\left(Y_q'\right){}^2\right){}^2}{\left(Y_d'\right){}^3-2
   \left(Y_q'\right){}^3+\left(Y_u'\right){}^3}+\frac{2}{3} \left(2 Y_d'-Y_q'\right)-\frac{5 Y_u'}{3} \label{eq:AllqYTAR} 
\end{dmath}
\begin{dmath}
Y'_{t} = \frac{-2 Y_u' \left(\left(Y_d'\right){}^2+\left(Y_q'\right){}^2\right)-4 \left(Y_d'\right){}^3+8
   \left(Y_q'\right){}^3}{\left(Y_d'\right){}^2+\left(Y_q'\right){}^2-2 \left(Y_u'\right){}^2} \label{eq:AllqYTR} 
\end{dmath}
\begin{dmath}
Y'_{b} = \frac{-2 Y_d' \left(\left(Y_q'\right){}^2-2 \left(Y_u'\right){}^2\right)-4 \left(Y_q'\right){}^3+2
   \left(Y_u'\right){}^3}{\left(Y_d'\right){}^2+\left(Y_q'\right){}^2-2 \left(Y_u'\right){}^2} \label{eq:AllqYBR} 
\end{dmath}
\begin{dmath}
Y'_{\mu} = \frac{-2 Y_d' \left(\left(Y_q'\right){}^2-2 \left(Y_u'\right){}^2\right)-4 \left(Y_q'\right){}^3+2
   \left(Y_u'\right){}^3}{\left(Y_d'\right){}^2+\left(Y_q'\right){}^2-2 \left(Y_u'\right){}^2} \label{eq:AllqYMU} 
 \end{dmath}
 \end{dgroup}
 
This construction is always possible as long as $\left(Y_d'\right){}^2+\left(Y_q'\right){}^2-2 \left(Y_u'\right){}^2 \neq 0$ and $\left(Y_d'\right){}^2-2 \left(Y_q'\right){}^2 + \left(Y_u'\right){}^2 \neq 0$
 
Of course with these conditions, there will be couplings to the first generation of quarks. However, the $\Delta M_s$ bound in models with a right-handed quark coupling is different than for those models discussed previously, since it is possible to obtain a negative contribution to $\Delta M_s$. Such a negative contribution can bring $\Delta M_s$ closer to the experimentally measured value than the current SM prediction, as noted in \cite{DiLuzio:2017fdq}, but we leave an exploration of this issue to a future study.

\end{document}